\documentclass[pre,amsfonts,floatfix,superscriptaddress,twocolumn,longbibliography]{revtex4-1}
\usepackage{mathbbol}
\usepackage{graphicx,color}
\usepackage{amsmath,amssymb,bm,amsthm}
\usepackage{mathtools}
\usepackage{simplewick}
\usepackage{braket}
\usepackage{mhchem}

\usepackage{mathrsfs}

\def\be{\begin{equation}}
\def\ee{\end{equation}}
\def\ba{\begin{eqnarray}}
\def\ea{\end{eqnarray}}

\begin{document}
\title{Level Repulsion and Dynamics in the Finite One-Dimensional Anderson Model}

\author{E. Jonathan Torres-Herrera}
\affiliation{Instituto de F{\'i}sica, Benem{\'e}rita Universidad Aut{\'o}noma de Puebla,
Apt. Postal J-48, Puebla, 72570, Mexico}
\author{J. A. M{\'e}ndez-Berm{\'u}dez}
\affiliation{Instituto de F{\'i}sica, Benem{\'e}rita Universidad Aut{\'o}noma de Puebla,
Apt. Postal J-48, Puebla, 72570, Mexico}
\author{Lea F. Santos}
\affiliation{Department of Physics, Yeshiva University, New York City, NY, 10016, USA}

\date{\today}

\begin{abstract}
This work shows that dynamical features typical of full random matrices can be observed also in the simple finite one-dimensional (1D) noninteracting Anderson model with nearest neighbor couplings. In the thermodynamic limit, all eigenstates of this model are exponentially localized in configuration space for any infinitesimal onsite disorder strength $W$. But this is not the case when the model is finite and the localization length is larger than the system size $L$, which is a picture that can be experimentally investigated. We analyze the degree of energy-level repulsion, the structure of the eigenstates, and the time evolution of the finite 1D Anderson model as a function of the parameter $\xi \propto (W^2 L)^{-1}$. As $\xi$ increases, all energy-level statistics typical of random matrix theory are observed. The statistics are reflected in the corresponding eigenstates and also in the dynamics. We show that the probability in time to find a particle initially placed on the first site of an open chain decays as fast as in full random matrices and much faster that when the particle is initially placed far from the edges. We also see that at long times, the presence of energy-level repulsion manifests in the form of the correlation hole. In addition, our results demonstrate that the hole is not exclusive to random matrix statistics, but emerges also for $W=0$, when it is in fact deeper.
\end{abstract}

\maketitle

\section{Introduction}

Anderson localization refers to the exponential localization in configuration space of the eigenstates of noninteracting systems with onsite disorder. It is caused by quantum interferences and was first studied in the context of electron conductance~\cite{Anderson1958,Lee1985,Kramer1993,Lagendijk2009,Abrahams2001}. The model employed to explain this phenomenon is a tight-binding model with random diagonal elements. In three dimensions (3D), the model may be found in the delocalized (metallic) or localized (insulating) phase, depending on the disorder strength, and a mobility edge is present.

The metal-insulator transition in the 3D model was also studied in terms of energy-level statistics~\cite{Efetov1983,Altshuler1988,Shklovskii1993,Mirlin2000}. The focus in~\cite{Shklovskii1993} was on the distribution $P(s)$ of the spacings $s$ between unfolded neighboring levels. The claim was that in the thermodynamic limit, there should be only three possible forms for $P(s)$: the Wigner-Dyson distribution [$P_{\text{WD}}(s)$] in the metallic phase, a Poissonian distribution [$P_{\text{P}}(s)$] in the localized phase, and an intermediate distribution between the two shapes at the critical point. 

The Poissonian distribution appears when the eigenvalues are uncorrelated and degeneracies are not prohibited. The Wigner-Dyson distribution, first studied in random matrix theory (RMT)~\cite{MehtaBook,Guhr1998}, emerges when the eigenvalues are highly correlated in the sense they strongly repel each other. In the increasing order of the degree of correlations, we find first the Gaussian orthogonal ensemble (GOE), when the full random matrices are real and symmetric, then the Gaussian unitary ensemble (GUE), when the full random matrices are Hermitian, and finally the Gaussian symplectic ensemble (GSE), when the full random matrices are written in terms of quaternions. In the case of the 3D Anderson model, the only $P_{WD}(s)$ observed is that for the GOE.

For the 1D Anderson model in the thermodynamic limit, all eigenstates are exponentially localized in configuration space for any infinitesimal value of the disorder strength, so there is no metal-insulator transition. However, if the 1D system is finite, we observe the crossover between different kinds of energy-level statistics. Instead of going only from $P_P(s)$ to $P_{WD}^{GOE}(s)$~\cite{Wischmann1990}, even stronger levels of correlations are seen. Indeed we have the following picture: The eigenstates are localized on the scale of the sample size if the disorder is strong, in which case the energy-level spacing distribution is Poissonian. In the other extreme, where disorder is absent, the eigenstates are extended and the eigenvalues are nearly equidistant. In this last case, the spectrum is said to be of the ``picket-fence'' type~\cite{Berry1977,Pandey1991}, which means that the eigenvalues are even more correlated than what one finds in full random matrices and the energy-level spacing distribution has an abrupt peak away from $s=0$. Between these two limits, as the disorder strength decreases from the point where the energy levels are uncorrelated [$P_{\text{P}}(s)$] to the point where they are picket-fence like, the spectrum experiences all degrees of energy-level repulsion typical of full random matrices, going first from Poisson to GOE-like, then from GOE-like to GUE-like, from GUE-like to GSE-like, and finally from GSE-like to picket-fence, passing through all intermediate distributions in between~\cite{Sorathia2012}. Notice, however, that these different energy-level spacing distributions do not reflect the symmetries of the system. From large disorder to zero disorder, the Hamiltonian matrix of the 1D Anderson model is throughout real and symmetric, just as GOE matrices. The different shapes of $P(s)$ are caused by the unavoidable change in the degree of correlations that the eigenvalues have to experience as going from being uncorrelated to picket-fence like.

Since experiments are done with finite systems~\cite{Katsumoto1987,Waffenschmidt1999,Billy2008,Roati2008,Manai2015,Volchkov2018}, we decided to look into the finite 1D Anderson model in more detail, focusing not only on energy-level statistics, but also on the structure of the eigenstates and on dynamics. We consider a chain with open boundary conditions. The analysis of the dynamics is particularly relevant to experiments with cold atoms and ion traps that have limited access to the spectrum, but routinely investigate time evolutions. Our studies are made with respect to the parameter $\xi \propto (W^2 L)^{-1}$, where $W$ is the disorder strength and $L$ is the size of the chain.

We find clear relationships between different energy-level repulsion parameters and measures of the degree of delocalization of the eigenstates. However, despite the similarities between the energy-level statistics of the finite Anderson model and RMT, not all eigenstates of the 1D system are random vectors as in full random matrices, since their structures depend on their energies.

To study the dynamics, we place the particle initially on a single site and compute the evolution of the probability to find it on a specific site of the chain. The survival probability, which is the
probability to find the particle on its initial state, receives special attention. Its initial decay cannot distinguish systems with or without level-repulsion, a claim that has been made in the past in the context of many-body quantum systems~\cite{Torres2014PRA,Torres2014NJP,Torres2014PRAb}. Manifestations of the correlations between the eigenvalues emerge only at long times, in the form of the correlation hole~\cite{Leviandier1986,Guhr1990,Wilkie1991,Alhassid1992,Gorin2002,Torres2017,Torres2017Philo,Torres2018,TorresHerrera2019,Sergio2019PRE}. The hole is not necessarily a signature of quantum chaos. As we show, it emerges also in the limit where $W=0$ and the model is integrable, but the eigenvalues are strongly correlated. The hole gets deeper as the correlations become stronger and the statistics moves from GOE-like to GUE-like, to GSE-like, and finally reaches the picket-fence spectrum.

Before the correlation hole, the evolution of the survival probability depends on the shape and bounds of the energy distribution of the initial state~\cite{Tavora2016,Tavora2017}. We find that when the particle is on the first site of the chain, the shape of this distribution is the same found for the dynamics under full random matrices, resulting in the same power-law decay $\propto t^{-3}$ observed for RMT~\cite{TorresKollmar2015,Tavora2016,Tavora2017}. This behavior is much faster than the $t^{-1}$ decay obtained for a particle initially placed far from the edges of the chain. The physical justification of this behavior lies in the spread of the initial state over other site-basis vectors, which is almost unidirectional for the particle on site 1, but more homogeneous for the particle in the middle of the chain. This is an interesting border effect that may find applications for fast and efficient transfers of particles in quantum systems.

The paper is organized as follows. In Sec.~II, we describe the model. Section~III is dedicated to the analysis of the spectrum and the structure of the eigenstates. The dynamics is the subject of Sec.~IV. 
Discussions are presented in Sec.~V. 

\section{Finite 1D Anderson Model}

The 1D Anderson model with onsite disorder and open boundaries is described by the one-particle tight-binding Hamiltonian,
\be
H = \sum_{n=1}^{L} \epsilon_n c_n^{\dagger} c_n - J \sum_{n=1}^{L-1} (c_n^{\dagger} c_{n+1} + c_{n+1}^{\dagger} c_{n} ).
\label{Eq:Anderson}
\ee
Above, $c_n^{\dagger}$ $(c_n)$ is the creation (annihilation) operator of a particle on site $n$,  $\epsilon_n$ are uniform random numbers in the interval $[-W/2,W/2]$, so their variance is $W^2/12$,   and $L$ is the finite number of sites. The energy scale is set by choosing the hopping amplitude $J=1$. 
The density of states (DOS) is bounded by the spectrum edges $\pm (2+W/2)$.

We use as basis vectors $|\phi_n \rangle$, states that have the single particle placed on a single site $n$. In this basis, the Hamiltonian matrix for Eq.~(\ref{Eq:Anderson}) is tridiagonal.
If the disorder strength $W$ is very strong, $\langle |\epsilon_n - \epsilon_m|\rangle \gg 1$, the diagonal part dominates and the eigenstates are localized in configuration space. This happens when the localization length is smaller than the chain size $L$. 

In the absence of disorder, $W=0$, we have a Toeplitz tridiagonal matrix. The eigenvalues are 
\be
E_\alpha = 2\cos \left( \frac{\alpha \pi}{L+1} \right), \quad \alpha=1,2,\ldots L
\label{Eq:E}
\ee
and the eigenstates are 
\be
|\psi_{\alpha} \rangle = \sum_{n=1}^L \sqrt{ \frac{2}{L+1} } \sin \left( \frac{\alpha n \pi}{L+1} \right) |\phi_n \rangle.
\label{Eq:psiAlpha}
\ee

We use $\alpha$ as the index for the eigenstates of $H$ and $n$ for the basis vectors $|\phi_n \rangle$. We denote the overlaps between $|\psi_{\alpha} \rangle $ and $|\phi_n \rangle$ as $C^{(n)}_{\alpha} = \langle \psi_{\alpha}  | \phi_n \rangle  = \left(C^{(n)}_{\alpha}\right)^* = \langle \phi_n |\psi_{\alpha} \rangle $.

\subsection{Parameter for the Analysis}

The localization length $l_{\infty} (E)$ for the 1D Anderson model was obtained by Thouless via perturbation theory to second order in the disorder strength~\cite{Thouless1974,ThoulessBook,Heinrichs2003},
\be
l_{\infty} (E) = \frac{96(1-E^2/4) }{W^2}.
\label{Eq:loc}
\ee
This expression is valid inside the energy band, not at the edges. It also has a correction right at the center of the spectrum~\cite{Czycholl1981,Kappus1981,Izrailev1998,Tessieri2012}, where it becomes
\be
l_{\infty} (0) =  \left[ \frac{\Gamma(1/4)}{\Gamma(3/4)} \right]^2 \frac{12  }{W^2} \approx \frac{105.045 }{W^2} .
\ee

Motivated by the above equations, our studies are carried out as a function of the parameter
\be
\xi=\frac{l_\infty(0)}{ L} \approx \frac{105.045 }{W^2 L},
\label{Eq:xi}
\ee 
which is varied from very small values, where for a fixed $L$, the disorder is very strong and the eigenstates are localized, to very large values, where the chain is almost clean and the eigenstates are maximally extended. Even though we vary $\xi$ over a large range of values, it is important to reiterate that Eq.~(\ref{Eq:loc}) holds only for weak disorder.

\section{Properties of the Eigenvalues and Eigenstates}
We analyze the properties of the eigenvalues and eigenstates as a function of the parameter $\xi$. Let us start with the DOS. In the clean limit, $W=0$, the DOS is obtained from~\cite{Schiulaz2018}
\ba
\rho_{\text{DOS}}(E) &=& \frac{1}{L} \sum_{\alpha} \delta(E - E_\alpha) \nonumber \\
&=& \frac{1}{L}\sum_{\alpha} \int_{-\infty}^{\infty} \frac{d\tau}{2\pi} e^{i \tau [E-2 \cos \left( {\alpha \pi}/{(L+1)} \right)]}.
\label{Eq:preDOS}
\ea
In the continuum, 
\ba
\rho_{\text{DOS}}(E) 
&=&  \int_{-\infty}^{\infty} \frac{d\tau}{2\pi} e^{i \tau E} {\cal J}_0(2\tau)  \nonumber \\
&=&  \frac{1}{\pi \sqrt{4 - E^2}},
\label{Eq:DOS}
\ea
where ${\cal J}_0$ is the Bessel function of first kind. The DOS has a U-shape, where the distribution peaks at the edges of the spectrum. This form persists even when the disorder is strong and the energy-level spacing distribution is Poissonian, provided $L$ is sufficiently large (for example, when $\xi \sim 0.002$ and $L\sim 16\, 000$, as we use below).

\subsection{Energy-Level Statistics}

To study the degree of short-range correlations between the eigenvalues, we use the energy-level spacing distribution $P(s)$ and also the ratio $\tilde{r}_\alpha$ between neighboring spacings. To get $P(s)$ we need to unfold the spectrum, but to compute $\tilde{r}_\alpha$ unfolding is not needed.

The Poissonian distribution is  $P_{\text{P}}(s) = \exp(-s)$. In the form of the Wigner surmise, the Wigner-Dyson distribution is written as $P_{\text{WD}}(s)=a_{\beta} s^{\beta} \exp(- b_{\beta} s^2)$. The constants $a_{\beta}$ and $b_{\beta}$ can be found in Ref.~\cite{MehtaBook,Guhr1998}, while $\beta$ is shown in Table~\ref{Tab:1} for the different energy-level statistics obtained for ensembles of full random matrices.

The ratio $\tilde{r}_\alpha$ is defined as~\cite{Oganesyan2007,Atas2012},
\be
\tilde{r}_\alpha = \min \left( r_\alpha, \frac{1}{r_\alpha}\right), \quad \text{where} \quad r_\alpha=\frac{s_\alpha}{s_{\alpha-1}}
\ee
and $s_\alpha = E_{\alpha+1} - E_\alpha$ is the spacing between neighboring levels. The values of $\langle \tilde{r}\rangle $ obtained by averaging $\tilde{r}_\alpha$ are also shown in Table~\ref{Tab:1} for different energy-level statistics. 
\begin{table}[h]
  \centering
  \begin{tabular}{ | c | c | c| c|}
    \hline
     & $\beta$ & $\langle \tilde{r}\rangle $ &$P_R$ \\ \hline
    Poisson & 0 & $2\ln2 -1 \approx 0.39$ & $\approx 1$ \\ \hline
    GOE & 1 & $4-2\sqrt{3} \approx 0.54$ & $(L+2)/3$ \\ \hline
    GUE & 2 & $2\sqrt{3}/\pi -1/2 \approx 0.60$ & $(L+1)/2$ \\ \hline
    GSE & 4 & $(32/15) \sqrt{3}/\pi -1/2 \approx 0.68$ & $(2L+1)/3$ \\  \hline
    clean &$\gg4$  & $\approx 1$ & $2(L+1)/3$ \\ 
    \hline
  \end{tabular}
 \caption{Values of $\beta$ and $\langle \tilde{r}\rangle $ for different level statistics and the corresponding  values of the participation ratio. The results for the GOE, GUE and GSE rows refer to those for ensembles of full random matrices. The result for $P_R$ for the clean model is obtained using Eq.(3).}
  \label{Tab:1}
\end{table}

To study $\beta$ as a function of $\xi$, we extract $\beta$ from the expression suggested in~\cite{Izrailev1990},
\begin{equation}
P_\beta (s) = A \left( \frac{\pi s}{2} \right)^\beta 
\exp \left[ -\frac{1}{4} \beta \left( \frac{\pi s}{2} \right)^2 - 
\left( B s -  \frac{\beta}{4} \pi s \right)  \right] .
\label{Eq:beta}
\end{equation}
The parameters $A$ and $B$ are obtained from the normalization conditions
\begin{equation}
\int_{0}^{\infty} P_\beta (s)  ds =1\,\,\text{and}\,\, \int_{0}^{\infty} s P_\beta (s)  ds =1.
\end{equation}
The phenomenological expression (\ref{Eq:beta}) was written so that it reproduces for $\beta=1,2$, and $4$ the energy-level spacing distributions obtained from RMT for GOE, GUE, and GSE, respectively.
Variations of Eq.~(\ref{Eq:beta}) can be found in~\cite{Casati1991,Kottos1999,Sorathia2012}.
  
\begin{figure}[htb]
\includegraphics[width=\columnwidth]{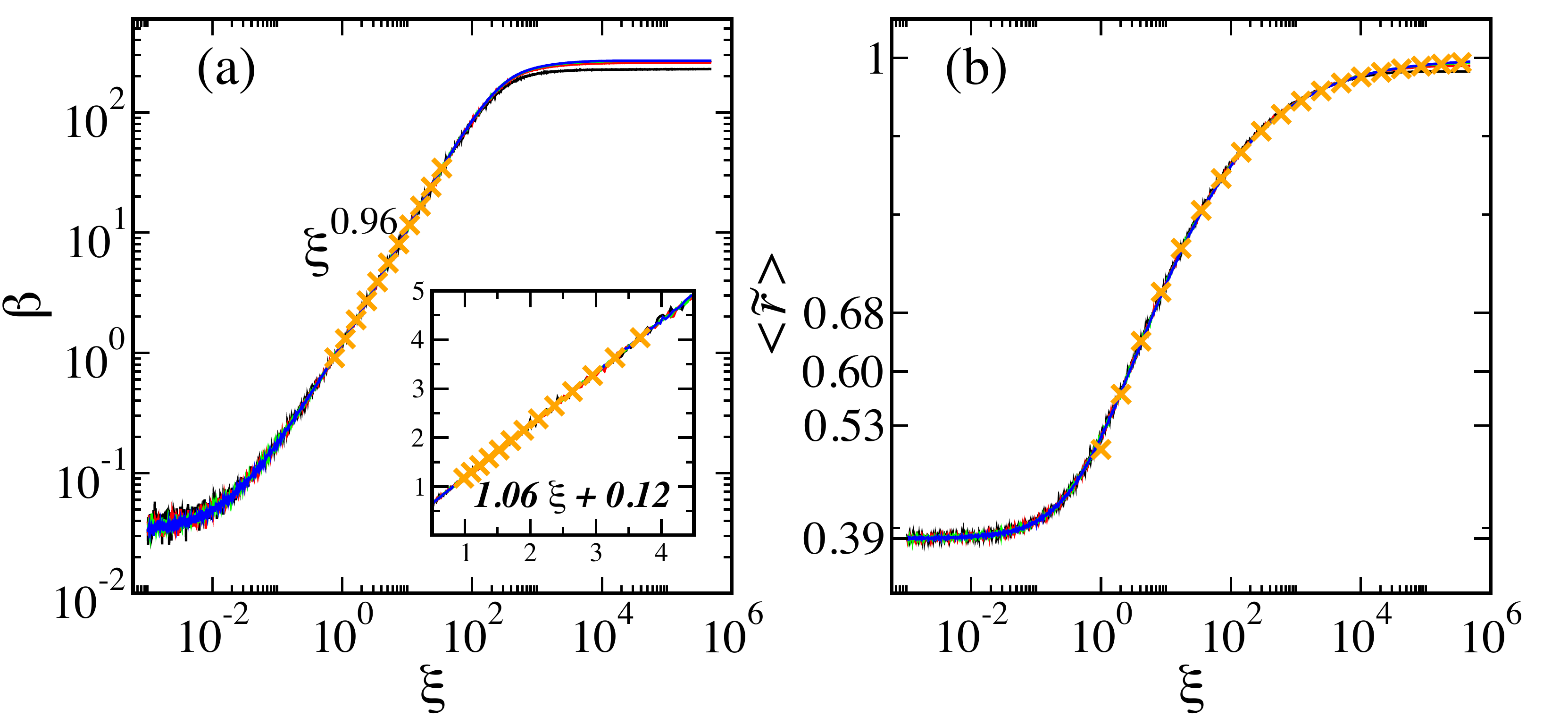}
\caption{Level repulsion parameter $\beta$ (a) and average ratio of spacings between consecutive levels $\langle \tilde{r}\rangle$ as a function of $\xi$. Four system sizes are considered, $L=512, 1024, 2048, 4096$, which are only distinguishable for very lager $\xi$ (hard to see for the chosen scales) (the size increases from bottom to top). Crosses indicate fitting curves: the linear functions are written in the main panel and inset of Fig.~\ref{fig:beta}~(a), while the function for Fig.~\ref{fig:beta}~(b) is given by Eq.~(\ref{Eq:sigmoid}). Averages over $100$ disorder realizations.}
\label{fig:beta}
\end{figure}

The main panel of Fig.~\ref{fig:beta}~(a) shows that $\beta$ is a linear function of $\xi$ for a broad range of values, $0.5 \lesssim  \xi \lesssim 10^2$. The linear fit becomes even better for the values of $\xi$ where we have the standard RMT distributions, that is for $1\leq \beta \leq 4$, as seen in the inset of Fig.~\ref{fig:beta}~(a).
This inset suggests that the energy-level repulsion parameter $\beta$ equals the rescaled localization length obtained at the middle of the spectrum~\cite{footnoteSorathia},
\begin{equation}
\beta \approx \xi. 
\label{Eq:linear}
\end{equation}
Following discussions in Ref.~\cite{Sorathia2012}, the linear relation $\beta \propto \xi$ appears to be quite general and has been observed also for the kicked rotor~\cite{Izrailev1989,Izrailev1990} and Wigner banded random matrices~\cite{Casati1993}; see also~\cite{Angel2019} where a linear relation between the repulsion parameter $\beta$ and the degree of disorder of the 1D Anderson model $\xi$ was obtained through scattering quantities.

Figure~\ref{fig:beta}~(b) shows $\langle \tilde{r}\rangle $ as a function of $\xi$. The average ratio of level spacings goes from $\langle \tilde{r}\rangle \approx 0.39$ (Poisson) for very small $\xi$ to $\langle \tilde{r}\rangle \approx 1$ (picket-fence) for very large $\xi$. Four system sizes are shown and the curves fall on top of each other for most values of $\xi$, except for very larger $\xi$ (hard to see for the chosen scales), already in the picket-fence region. In this case, $\langle \tilde{r}\rangle \rightarrow 1$ as $L$ increases, as it should be, since the contributions from the edges of the spectrum, where the neighboring eigenvalues are not equidistant, become negligible. At the edges of the spectrum, according to Eq.~(\ref{Eq:E}), we have $r_{2} =5/3$ and $r_{L-1} =3/5$, while in the middle, $r_{(L+1)/2} =1$. 

Contrary to $\beta$, we see in Fig.~\ref{fig:beta}~(b) that $\langle \tilde{r}\rangle$ is not a linear function of $\xi$. Its behavior is well fitted with the following function,
\begin{equation}
\langle \tilde{r}\rangle \approx \frac{\xi^{0.4} }{1 + \xi^{0.4}} 
 \quad \text{for} \quad \xi > 1.
 \label{Eq:sigmoid}
\end{equation}
Taking Eq.~(\ref{Eq:linear}) into account, an expression very similar to Eq.~(\ref{Eq:sigmoid}) describes $\langle \tilde{r}\rangle$ {\em vs} $\beta$ for $1\leq \beta \leq 4$. 

Notice that the values of $\xi$ that we obtain for each of the RMT level statistics using Eq.~(\ref{Eq:sigmoid}) are in the vicinity of what we get using Eq.~(\ref{Eq:linear}), but they do not agree exactly. For the GOE case, $\langle \tilde{r} \rangle_{GOE} = 0.54$ leads to $\xi \sim 1.5$, instead of $\xi \approx \beta =1$. The discrepancy increases as the correlations get stronger. For $\langle \tilde{r} \rangle_{GSE} = 0.68$, we reach $\xi \sim 6.6$, instead of $\xi \approx \beta =4$. The disagreement may in part be due to numerical factors, such as the unfolding procedure and the relatively small chain sizes, but also to the fact that we are not dealing with full random matrices and level repulsion caused by symmetries, but instead with a realistic model with short-range couplings described by real and symmetric matrices.

\subsection{Structure of the States}
The degree of localization of the eigenstates has been studied with measures, such as the localization length~\cite{Casati1992} and entropic quantities like the so-called structural entropy~\cite{Varga1994}. Here, we provide a detailed analysis of the structure of eigenstates and contrast our results with those for full random matrices.

To analyze the eigenstates, we use the participation ratio,
\begin{equation}
P_{R}^{\alpha} = \frac{1}{\sum_{n=1}^L|C^{(n)}_{\alpha}|^4 },
\end{equation}
which quantifies the number of basis vectors $|\phi_n \rangle $ that contribute to the eigenstate $|\psi_\alpha \rangle$. According to Table~\ref{Tab:1}, for full random matrices, in the region $1\leq \beta \leq 4$, we have approximately
\be
\frac{\langle {P_{R}^\alpha}^{RMT}\rangle}{L}  \sim \frac{\sqrt{\beta} }{3}.
\label{Eq:prRMT}
\ee

\begin{figure}[htb]
\includegraphics[width=\columnwidth]{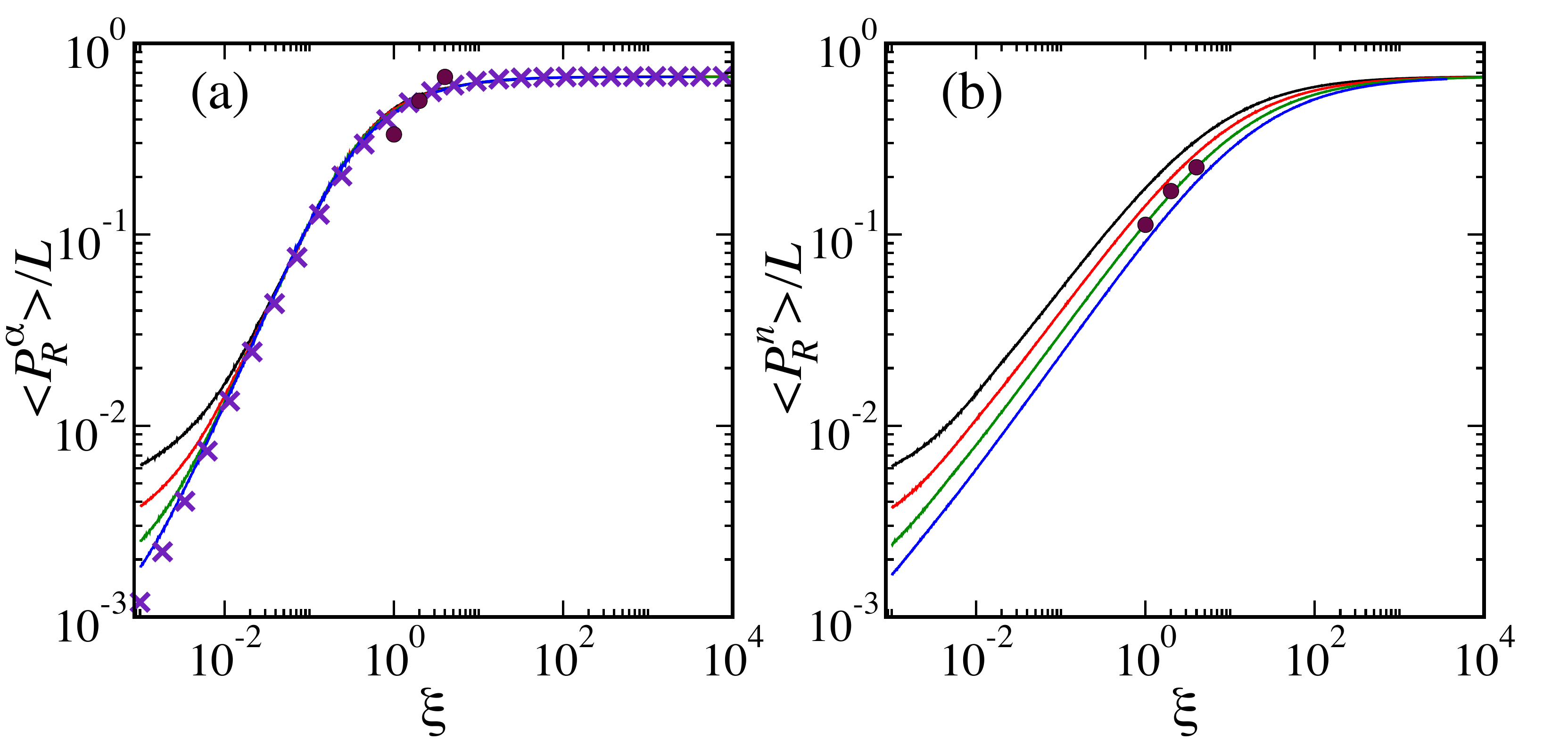}
\caption{Scaled average participation ratio for eigenstates in the middle of the spectrum (a) and for basis vectors (b) as a function of $\xi$. System sizes are (from top to bottom) $L=256,\,512,\,1024,\,2048$. Circles in (a) are the RMT predictions for $\beta=1,\,2,\,4$, while crosses (orange) represent Eq.~\eqref{eq:PRvsLRM}.  Circles in (b) are obtained from Eq.~\eqref{eq:PRvsLAM} with $\kappa=1/3$. For the averages we collected $10^4$ statistical data obtained from disorder realizations and states around to the middle of the spectrum.}
\label{fig:PR}
\end{figure}

In Fig.~\ref{fig:PR}~(a), we report the average value of $P_{R}^{\alpha}$ for the eigenstates $|\psi_{\alpha}\rangle$ of the Anderson model at the middle of  the spectrum as a function of $\xi$.
The results make it evident that a relationship exists between the degree of delocalization of the eigenstates and the degree of level repulsion (recall that $\xi \approx \beta$), but it also exposes the limits of the analogy between the properties of full random matrices and of the finite 1D Anderson model. For the Anderson model, in the region $1\leq \beta \leq 4$, we actually find
\be
\frac{\langle P_{R}^{\alpha} \rangle}{L}  \sim \frac{0.67 \beta}{0.56 + \beta},
\label{eq:PRvsLRM}
\ee
instead of something closer to Eq.~(\ref{Eq:prRMT}). Indeed, the circles in the figure, which indicate the values from RMT,  do not match exactly the values of $\langle P_{R}^{\alpha} \rangle/L$ for $\beta=1$, 2, and 4. 

The differences between the realistic model and RMT are not only restricted to the average values of the participation ratio, but extends also to their energy dependence. Contrary to what we find in RMT, where all eigenstates are random vectors, so $\langle P_{R}^{\alpha} \rangle$ {\em vs} $\langle E_{\alpha} \rangle$ is a flat curve, for the Anderson model, the structure of the eigenstates depends on the energy. As it is known, they are more extended close to the middle of the spectrum than at the edges. 

In Fig.~\ref{fig:PR}~(b), we depict the degree of delocalization of the basis vectors $|\phi_n\rangle$ projected into the energy eigenbasis, that is $\langle P_{R}^{n} \rangle/L$ {\em vs} $\xi$. As we will see in the next section,  $|\phi_n\rangle$ are the initial states that we consider for the analysis of the dynamics. Interestingly, the dependence of this participation ratio on $\beta$ (recall that $\xi \approx \beta$) is similar to that for RMT, apart from a factor $\kappa$ that varies with system size. We find that in the region where $1\leq \beta \leq 4$ (but not beyond that),
\be
\frac{\langle P_{R}^{n}\rangle}{L}  \sim \kappa \frac{\sqrt{\beta} }{3}.
\label{eq:PRvsLAM}
\ee
For $L=1024$, we have $\kappa \sim 1/3$, and this factor decreases as $L$ increases, suggesting that the basis vectors written in the energy eigenbasis must be multifractal~\cite{Schreiber1991}. This separation between the curves in Fig.~\ref{fig:PR}~(b) contrasts with Fig.~\ref{fig:PR}~(a), where the curves for $\langle P_{R}^{\alpha} \rangle/L$ for different chain sizes and $\xi > 10^{-2}$ fall on top of each other. In the case of $\langle P_{R}^{\alpha} \rangle$, the curves do not overlap only in the insulating phase, where the eigenstates become localized. 

In Fig.~\ref{fig:PR} the fluctuations in the values of $P_{R}/L$ are small, the largest dispersions being $3\%$ of the average $\langle P_{R}\rangle/L$.

After the characterization of the model in terms of level statistics and of the structure of the eigenstates, we now proceed with the analysis of the dynamics.

\section{Time Evolution}

To study the time evolution of the model, we assume that the particle is initially on the site $n_0$, so $ | \Psi(0) \rangle=|\phi_{n_0}\rangle$. This state, of energy $\epsilon_{n_0}$, evolves as
$|\Psi(t)\rangle = e^{-iHt} | \Psi(0) \rangle$. 

\subsection{Survival Probability}

We start by analyzing the survival probability in time, which is the probability to find the particle in its initial position at time $t$. It is given by
\ba
\label{Eq:SP}
S_{P_{n_0}} (t)&=&\left| \langle \Psi(0) | e^{-iHt} | \Psi(0) \rangle \right|^2 \nonumber \\
&=& \left| \sum_{\alpha}  \left|C_\alpha^{(n_0)}\right|^2 e^{-iE_\alpha t} \right|^2  \nonumber \\
&=& \left| \int_{E_{low}}^{E_{up}} \rho_{n_0}(E) e^{-iEt} dE \right|^2  .
\ea
Above,  
\be
\rho_{n_0}(E)=\sum_{\alpha} \left| C_\alpha^{(n_0)} \right|^2 \delta(E-E_{\alpha})
\label{Eq:LDOS}
\ee
is the energy distribution of the initial state, which is also known as local density of states (LDOS) or strength function, and
$E_{low}$ and $E_{up}$ are the energy bounds of the LDOS (in our case, $E_{low}=-2-W/2$ and $E_{up}=2+W/2$). The variance $\sigma_{n_0}^2$ of the LDOS depends on the number of states directly coupled with the initial state,
\ba
\sigma_{n_0}^2 &=&  \langle \Psi(0) |H^2| \Psi(0) \rangle  -   \langle \Psi(0) |H| \Psi(0) \rangle ^2 \nonumber \\
&=&\sum_{n \neq n_0} |  \langle \phi_n |H| \phi_{n_0} \rangle |^2 .
\label{eq:Gamma}
\ea
Since the chain has open boundaries, $| \Psi(0) \rangle$ is directly coupled with $2$ states when $n_0\neq 1, L$ and with $1$ state when $n_0= 1, L$.
By knowing the LDOS in detail, we should be able to determine the behavior of the survival probability, since the later is the square of the Fourier transform of the former.

\subsubsection{Local Density of States and the Survival Probability for $W=0$}

To get insight on the behavior of the survival probability, we study first the LDOS  for the clean model ($W=0$). The shape of the LDOS depends on the site $n_0$, where we initially place the particle, as
\be
\rho_{n_0}(E)= \frac{2}{L+1}  \sum_{\alpha=1}^L \sin^2 \left( \frac{\alpha n_0 \pi}{L+1} \right) \delta(E-E_{\alpha}) .
\label{Eq:rhono}
\ee
The sum above can be solved exactly (see Appendix~\ref{sec:derivation}), which gives the LDOS in the following closed form,
\be
\rho_{n_0}(E)=\frac{1-T_{2n_0}\left(\frac{E}{2}\right)}{\pi\sqrt{4-E^2}} ,
\label{eq:LDOSAn}
\ee
where $T_{2n_0}$ are the Chebyshev polynomials of the first kind.

For $| \Psi(0) \rangle = |\phi_1\rangle $, in particular, the shape of the LDOS is semicircular,
\be
\rho_{1}(E)=\frac{1-T_{2}\left(\frac{E}{2}\right)}{\pi\sqrt{4-E^2}}=\frac{\sqrt{4 - E^2}}{2 \pi } .
\ee
which is the same shape of the LDOS that we obtain for states that evolve under full random matrices~\cite{Torres2014PRA,Torres2014NJP,Torres2014PRAb}. Some examples for other sites $n_0$, also derived from Eq.~\eqref{eq:LDOSAn}, are listed below,
\ba
\rho_{2}(E)&=& E^2 \rho_1(E) ,\nonumber \\
\rho_{3}(E)&=& (E^2-1)^2 \rho_1(E) ,\nonumber \\
\rho_{4}(E)&=& E^2(E^2-2)^2 \rho_1(E) .\nonumber
\ea
As we gradually move the particle from $n_0=1$ to the center of the chain, the number of peaks in the LDOS increases, this number actually coincides with $n_0$. The peaks get more prominent and concentrated at the edges of the spectrum. Once the particle is in the middle of the chain, $n_0 = (L+1)/2$, all overlaps between the initial state and the energy eigenbasis are equal, 
\ba
&&\rho_{(L+1)/2}(E) = \frac{2}{L+1}  \sum_{\alpha=1}^L \sin^2 \left( \frac{\alpha  \pi}{2} \right) \delta(E-E_{\alpha}) \nonumber \\
&&= \frac{1}{L+1}  \sum_{\alpha=1}^L  \delta(E-E_{\alpha}) 
=\frac{1}{\pi \sqrt{4 - E^2}},
\label{Eq:recover}
\ea
so we recover the U-shape obtained for the DOS, as given by  Eq.~(\ref{Eq:preDOS}) and Eq.~(\ref{Eq:DOS}). This result also follows from Eq.~\eqref{eq:LDOSAn} for $L\gg1$.

Since the shape of the LDOS determines the decay of the survival probability via the Fourier transform in Eq.~(\ref{Eq:SP}), using Eq.~\eqref{eq:LDOSAn} (see Appendix~\ref{sec:derivation}), we also obtain a closed form for the survival probability,
\be
S_{P_{n_0}}(t)=\left[{\cal J}_0 (2 t)-(-1)^{n_0}{\cal J}_{2n_0} (2 t)\right]^2.
\label{eq:SPno}
\ee
The equations for some particular values of $n_0$ are given below,
\ba
&& S_{P_1} (t) =\frac{[{\cal J}_1 (2  t)]^2 }{ t^2}, \nonumber \\
&& S_{P_2} (t) =  \frac{4[{\cal J}_2 (2 t) - 2 t {\cal J}_3(2t)]^2 }{ t^4} , \nonumber \\
&& S_{P_3} (t) = \frac{9 \left[ \left(3 t^3-20 t\right) {\cal J}_1(2 t)-4 \left(3 t^2-10\right) {\cal J}_2(2 t)\right]^2}{t^8} , \nonumber \\
&&\vdots  \nonumber \\
&&S_{P_{(L+1)/2}} (t) = [{\cal J}_0 (2  t)]^2.
\label{Eq:Bessel}
\ea

\subsubsection{Power-law Decay of the Survival Probability for $W\neq 0$}

\begin{figure}[htb]
\includegraphics[width=\columnwidth]{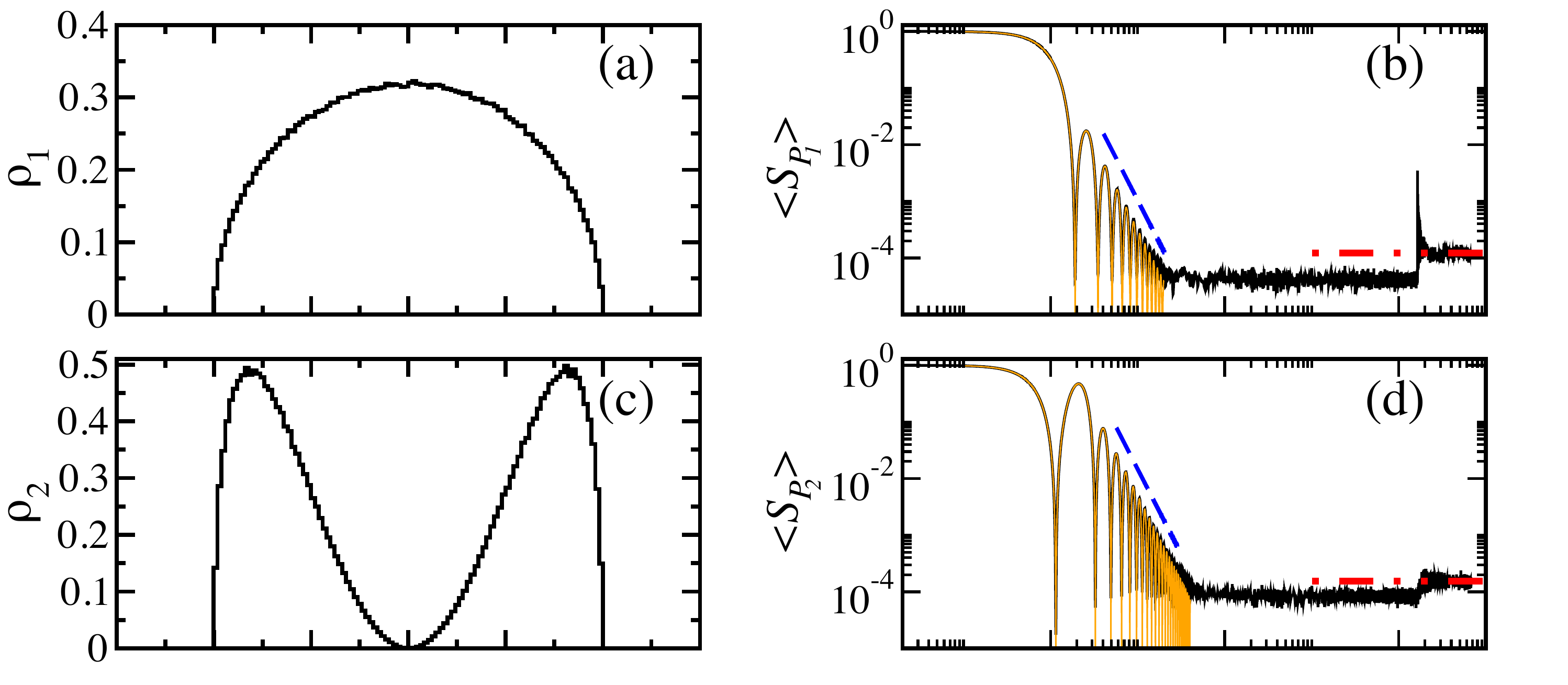}\\
\includegraphics[width=\columnwidth]{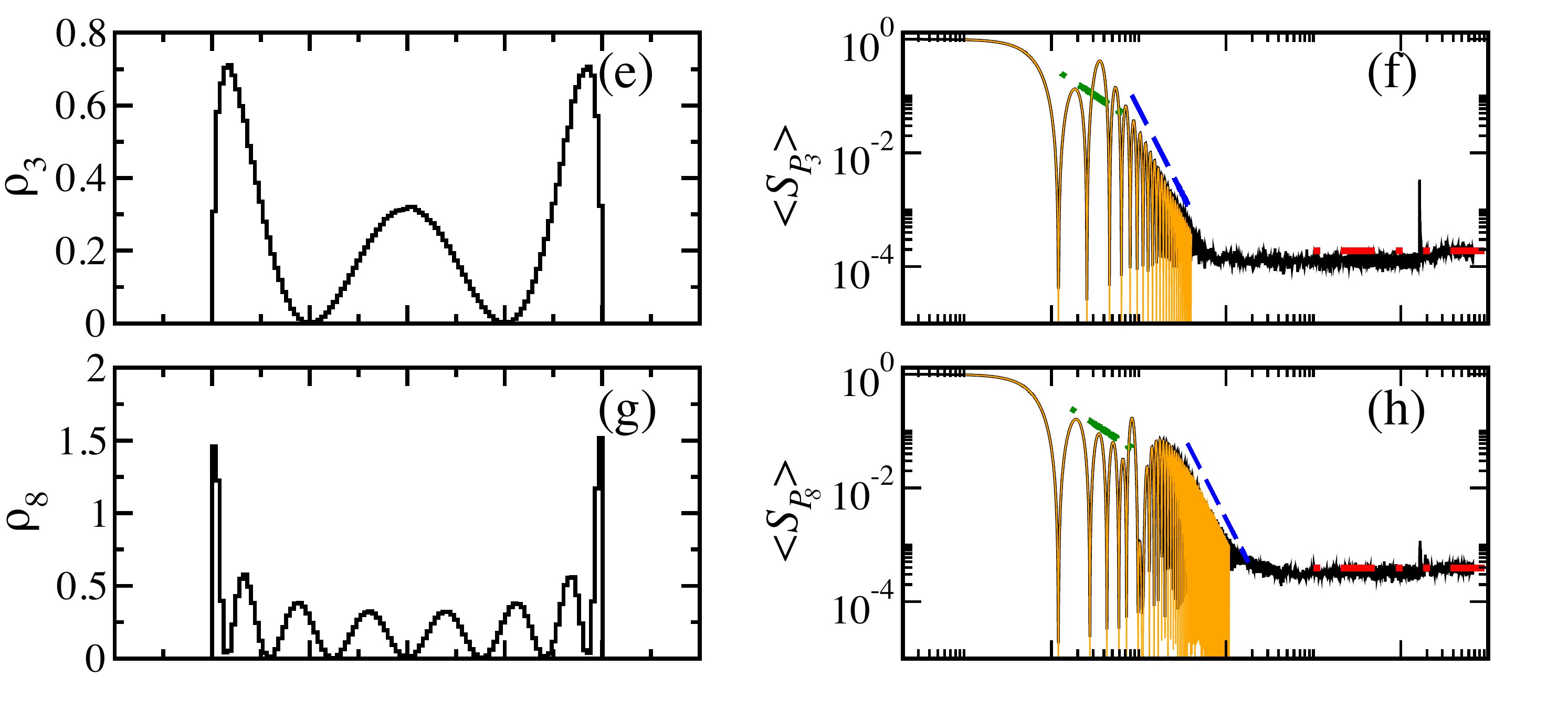}\\
\includegraphics[width=\columnwidth]{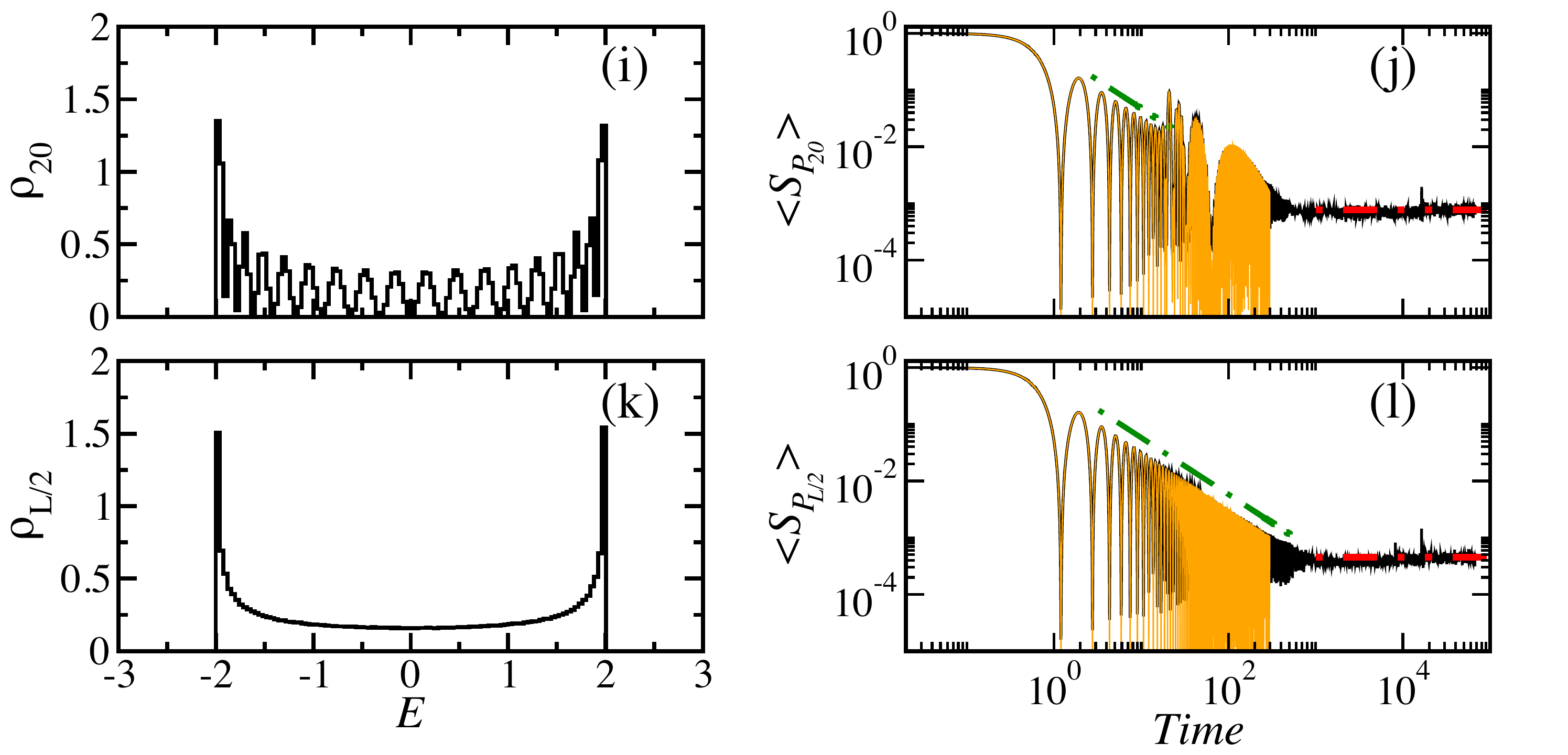}
\caption{Local density of states (left) and survival probability (right) for GSE-like statistics;  $n_0=1,2,3,8,20,L/2$ from top to bottom. Right panels: Dashed (Blue) and dot-dashed (green) lines correspond to the asymptotic decays $t^{-3}$ and $t^{-1}$, respectively. Horizontal double dotted-dashed (red) lines correspond to the infinite-time average.  In addition to the numerical curves for $W\neq0$, we show with orange lines the corresponding analytical expressions (\ref{Eq:Bessel}) for $W=0$. All panels are for $L=16\,384$; averages over $10^3$ disorder realizations; and $\xi=6.31$, which implies $\langle \tilde{r}\rangle \sim \langle \tilde{r} \rangle_{GSE}$.}
\label{fig:SP}
\end{figure}

Remarkably, when $W\neq 0$, the shapes of the LDOS remain very similar to those obtained for $W=0$ for values of $\xi \approx \beta \geq 1$, that is, for cases where the eigenvalues are strongly correlated and the localization length is larger than the system size. This is confirmed in the left panels of Fig.~\ref{fig:SP}, where the shapes obtained numerically with $W\neq0$ closely agree with those obtained from Eq.~(\ref{eq:LDOSAn}). In this figure, from the top  to the bottom panel, the initial position of the particle goes from site $n_0=1$ to site $n_0=L/2$. The results are shown for eigenvalues that have statistics close to GSE, but similar results are found for GOE and GUE.

The right panels of Fig.~\ref{fig:SP} display numerical results for the survival probability for $W \neq 0$ (black lines). For comparison, we also show the analytical expressions in Eq.~(\ref{Eq:Bessel}), obtained for $W=0$ (orange lines). The analytical curves are shown only up to $t=300$, since they go to zero for larger times, while numerical results saturate (see discussion about the saturation in the next subsection). Because the survival probability is not self-averaging~\cite{SchiulazARXIV2019}, in order to clearly reveal all of its dynamical features, we perform averages over ensembles of $10^3$ disorder realizations. The agreement between the analytical and numerical results is excellent. This confirms that the number of disorder realizations considered in the time interval $[0,300]$ is enough.

The behavior of $|\Psi(0)\rangle = |\phi_1 \rangle$ is rather counterintuitive. Only at very short times, the decay of $S_{P_1}(t)$ is slightly slower than for the other initial states, since the particle is attached to the border.
But for $t \gg 1$, the asymptotic behavior of $S_{P_1} (t)$ is $\propto t^{-3}$, as seen in Fig.~\ref{fig:SP}~(b). This power-law decay is what we have for evolutions under full random matrices~\cite{TorresKollmar2015,Tavora2016,Tavora2017}, where the LDOS is also semicircular. The evolution under full random matrices corresponds to the fastest dynamics that one can have for many-body systems with a unimodal LDOS~\cite{Torres2014PRAb}. It is  surprising to find the same behavior in a system with a single particle, and not only for $W=0$, but also in the presence of onsite disorder. 

As we change the initial position of the particle, placing it further from the edge, the $t^{-3}$ behavior gets postponed to later times, while the initial decay is substituted by the power-law $t^{-1}$, as seen in the right panels of Fig.~\ref{fig:SP}, from top to bottom. When $n_0$ is closer to the middle of the chain [Fig.~\ref{fig:SP}~(l)], only the $t^{-1}$ decay is seen~\cite{Arias1998,Santos2016}, which is the asymptotic behavior of $[{\cal J}_0 (2  t)]^2$ in Eq.~(\ref{Eq:Bessel}).

The power-law exponent of the algebraic decay of $S_{P}(t)$ can be derived from the bounds of the LDOS as follows~\cite{Erdlyi1956,Urbanowski2009,Nowakowski2008,Tavora2016,Tavora2017}. If $\rho_{n_0}(E)$ decays abruptly close to one of the energy bounds, such that 
$\rho_{n_0}(E)=(E-E_\text{low})^{\nu}\eta(E)$, for $0<\nu<1$ and $\lim_{E\to E_\text{low}}\eta(E)>0$, then $S_{P}(t)\propto t^{-2(\nu+1)}$. This explains the $t^{-3}$ decay, since for the semicircular LDOS, $\nu=1/2$. For the case where $n_0$ is away from the edges of the chain and the LDOS has a U-shape, we actually have $\nu = -1/2$, which is outside the range of values for $\nu$ above, but the relationship still holds, leading to the $t^{-1}$ decay.

\subsubsection{Correlation Hole}

The power-law decay is followed by the correlation hole, which is a dip below the saturation value of the survival probability. The saturation value is indicated with a horizontal double dotted-dashed line on the right panels of Fig.~\ref{fig:SP}. It corresponds to the infinite-time average of the survival probability,
\[
\overline{S_{P}}_{n_0} = \sum_{\alpha=1}^{L} |C_{\alpha}^{(n_0)}|^4 .
\]
The correlation hole gets below this point and develops only when the eigenvalues are correlated. It is nonexistent when the eigenvalues are uncorrelated and the $P(s)$ is Poissonian. 

As it is evident from Fig.~\ref{fig:SP}, the hole is better visible when the decay is entirely $\propto t^{-3}$ than when the decay is $\propto t^{-1}$, because in the former case, $S_{P_1} (t)$ has enough time to reach very low values.
One sees from the top panels that the survival probability is already below $\overline{S_{P}}_{n_0} $ for $t\gtrsim10$, which is not a very long time. This suggests that the correlation hole might be detectable experimentally even in platforms without exceedingly long coherence times. The same features observed in the figure are present also for $L\sim 100$, which are sizes within the reach of experiments with ion traps.

\begin{figure}[htb]
\includegraphics[width=\columnwidth]{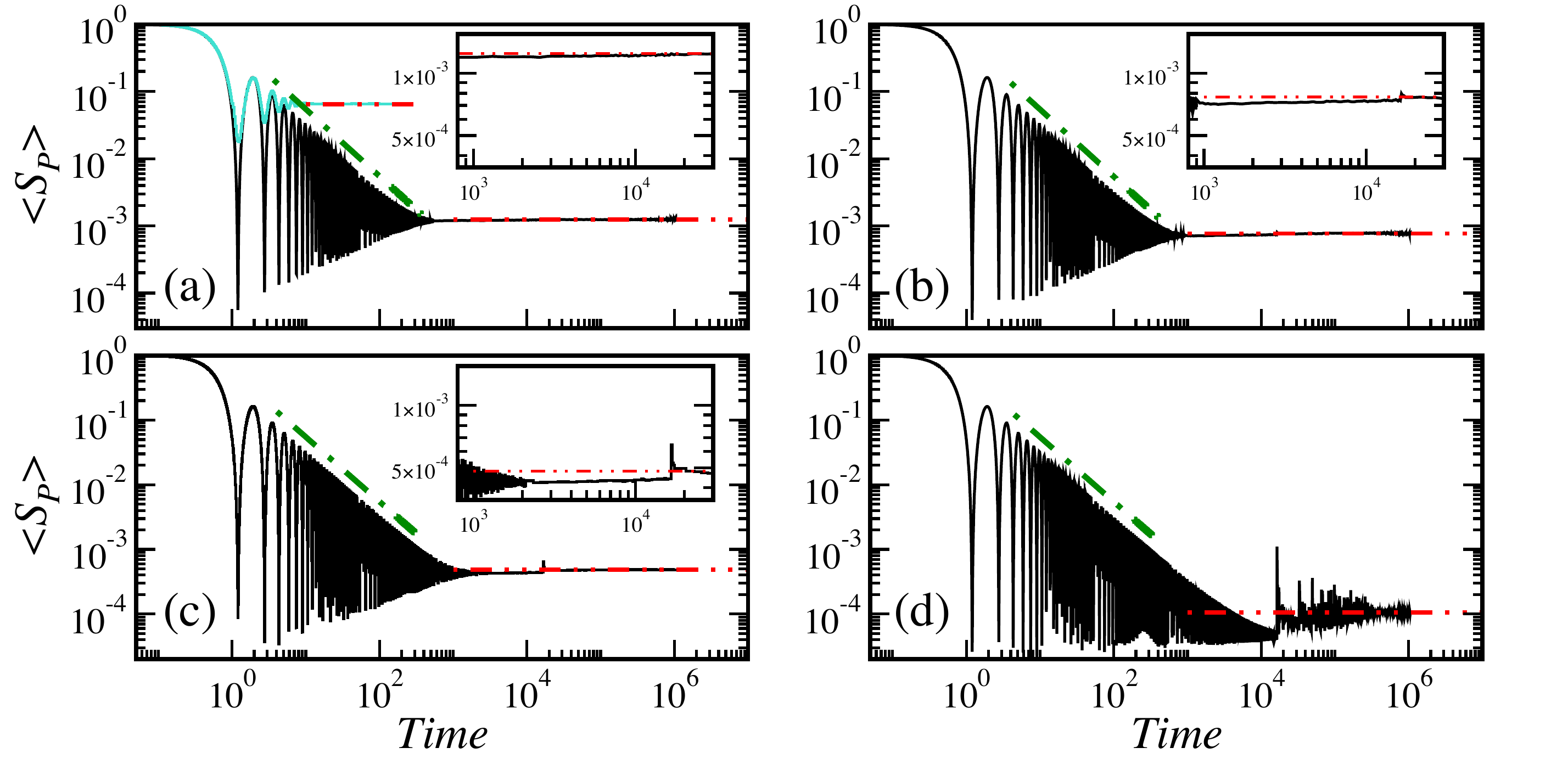}
\caption{Survival probability for initial states with $\epsilon_{n_0}\approx 0$ for Poissonian (turquoise) and GOE-like nearest-neighbor energy-level spacing distributions (a), GUE-like (b), GSE-like (c) and the picket-fence spectrum with $W\approx 0$ ($\xi=1000$)  (d).  Dotted-dashed lines indicate the $t^{-1}$ decay. Horizontal double dotted-dashed lines correspond to the infinite-time average. The insets in (a), (b) and (c) zoom in the correlation hole for the GOE-, GUE-, and GSE-like statistics. $L=16\,384$; averages over initial states and disorder realizations that give a total of $10^4$ data. An additional average is done over small windows of time to further reduce fluctuations.}
\label{fig:Hole}
\end{figure}

The correlation hole becomes deeper as the spectrum becomes more rigid (larger $\xi, \beta$), as seen in Fig.~\ref{fig:Hole}, where we go from Poisson and GOE-like [Fig.~\ref{fig:Hole}~(a)] to GUE-like [Fig.~\ref{fig:Hole}~(b)], GSE-like [Fig.~\ref{fig:Hole}~(c)], and picket-fence [Fig.~\ref{fig:Hole}~(d)]. In this figure, $n_0=L/2$. Notice that the hole is not exclusive to spectra that show statistics similar to those of RMT, but develops whenever correlations are present. The strongest correlations for the tight-binding model in Eq.~(\ref{Eq:Anderson}) happen for $W = 0$, when the eigenvalues away from the borders of the spectrum become nearly equidistant and the spectrum is of the picket-fence type. The hole is therefore deeper in Fig.~\ref{fig:Hole}~(d).

Figure~\ref{fig:Hole} also makes evident that the behavior of the survival probability before the correlation hole is independent of the level statistics. This observation has been made also in studies of nonequilibrium many-body quantum dynamics~\cite{Torres2014PRA,Torres2016Entropy}. In Fig.~\ref{fig:Hole}, since $n_0=L/2$ and the LDOS has the U-shape,  the same power-law decay $\propto t^{-1}$ emerges for all level spacing distributions, even for the Poissonian $P(s)$.  To capture correlations among the eigenvalues, times beyond the power-law decay need to be reached for the dynamics to resolve the discreteness of the spectrum~\cite{Schiulaz2019}. 

After the correlation hole, the dynamics finally saturates. Notice that in this region of equilibration, where the survival probability fluctuates around $\overline{S_{P}}_{n_0}$, one sees isolated peaks. They are visible in Fig.~\ref{fig:SP}, in the insets of Fig.~\ref{fig:Hole}, and specially in Fig.~\ref{fig:Hole} (d). They happen at times that are integer multiples of the system size $L$, when the particle reaches the edges of the chain and partial revivals take place.  Recurrences of this kind have been discussed in the context of a 1D quantum Ising chain at criticality~\cite{CamposVenuti2010}.  Experimental observations of recurrences in small quantum systems were verified in~\cite{Rempe1987,Greiner2002}.

\subsection{Particle Spread on the Lattice}

\begin{figure}[htb]
\includegraphics[scale=0.35]{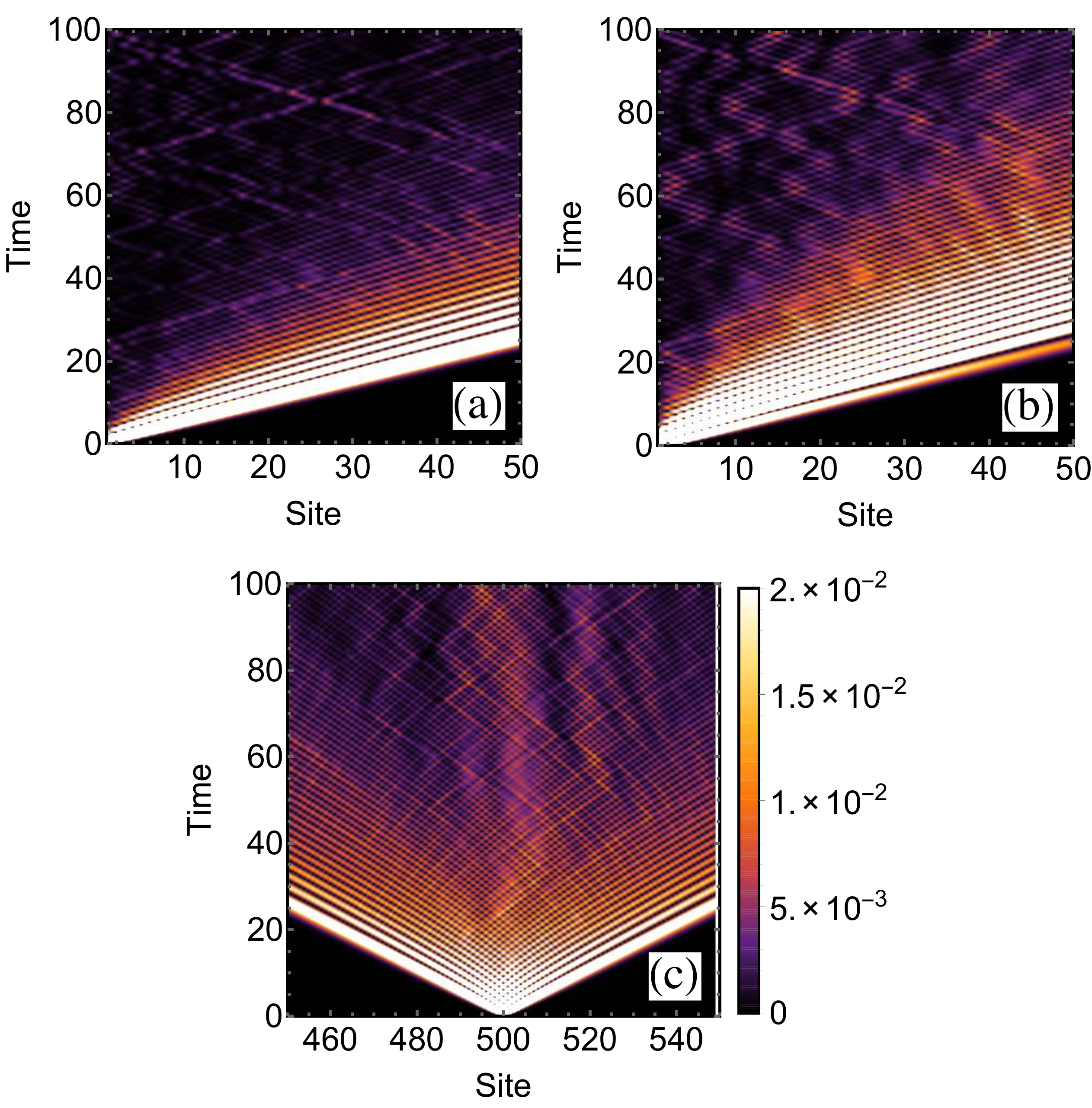}
\caption{Spread in time of the probability to find the particle on any site. Three initial states are shown, particle on $n_0=1$ (a), particle on $n_0=2$ (b), and particle on $n_0=L/2$ (c). Parameters: $\beta\approx 4$, $L=1000$, and one random realization.}
\label{fig:2D}
\end{figure}

To get a better picture of why the dynamics is faster when the particle is initially on the edge of the chain than when $n_0$ is away from the edges, we study how the probability to find the particle on any site spreads in time. This is shown with density plots in Fig.~\ref{fig:2D} for $n_0=1$ (a), $n_0=2$ (b), and $n_0=L/2$ (c). The checkerboard pattern of the panels are reminiscent of the Bessel functions that characterize the dynamics [see Eqs.~\eqref{Eq:Bessel}].

The motion of the particle for $|\Psi(0)\rangle = |\phi_1\rangle$ is very concentrated [Fig.~\ref{fig:2D}~(a)]. The particle moves away from the edge, hopping successively from site $n$ to site $n+1$, with very little spreading onto sites in the vicinity. For the case  $|\Psi(0)\rangle = |\phi_2\rangle$, the particle hops from site 2 to site 1 at $t\sim1$, and from there, its motion is again well directed [Fig.~\ref{fig:2D}~(b)], although there is more leaking than in Fig.~\ref{fig:2D}~(a). These behaviors contrast with that in Fig.~\ref{fig:2D}~(c), where the particle very soon spreads simultaneously through several sites in the vicinity of $n_0$ and has a significant chance to be brought back to its initial position even late in time. This is clearly not the case for Fig.~\ref{fig:2D}~(a). There, we see that once the particle detaches from the border, it is more difficult to bring it back to $n_0=1$. As a result, the particle moves away from $n_0=1$ faster than from $n_0=L/2$.

\section{Discussion}

All degrees of energy-level repulsion typical of full random matrices and even stronger (picket-fence like) can be obtained by varying the disorder strength and the system size of the 1D Anderson model, specifically by increasing $\xi \propto (W^2 L)^{-1}$ from very small to very large values. These different degrees of level correlations can be detected directly from the analysis of the spectrum or by studying the dynamics of the system. Both approaches were explored in this work.

Our analyses of the spectrum and also of the degree of delocalization of both the eigenstates and the initial states were done as a function of $\xi$. The parameter $\beta$, obtained from the level spacing distribution, depends linearly on $\xi$, while a less trivial relationship emerges for $\xi$ and the average ratio $\langle \tilde{r}\rangle $ between neighboring level spacings. The latter is a recently introduced quantity that has been gaining popularity in analysis of level repulsion. 

Correlated eigenvalues affect the dynamics of quantum systems by creating the so-called correlation hole. As we show, the correlation hole reflects all the different degrees of energy-level repulsion achieved by the 1D Anderson model, with stronger correlations leading to deeper holes. The correlation hole has been the subject of recent studies associated with many-body quantum systems in high energy physics~\cite{Cotler2017GUE,Gharibyan2018,Nosaka2018} and condensed matter physics~\cite{Torres2017Philo,Torres2018,Schiulaz2019,TorresHerrera2019}. However, for many-body quantum systems with local short-range couplings, the hole emerges at exceedingly long times~\cite{Schiulaz2019}, which prevents it from being observed experimentally. In this work, we showed that the correlation hole in the 1D Anderson model happens at times that are experimentally accessible.

To investigate the dynamics, the particle was initially placed on a single site. When this is the first site of the chain, or a site very close to the edges, the survival probability shows a power-law decay $\propto t^{-3}$, which is the same behavior one encounters when dealing with full random matrices. This fast decay allows for the emergence of the correlation hole at times that are not very large, being thus reachable by current experiments. In contrast, when the particle is initially close to the middle of the chain, the survival probability decays as $t^{-1}$ and the correlation hole takes longer to become visible.

The difference in the value of the power-law exponent for the survival probability decay is, of course, nonexistent in a chain with periodic boundaries, where the only power-law behavior observed is $\propto t^{-1}$. Yet, experimental setups usually deal with open chains. This border effect can be explored to better transfer particles and to detect dynamical features typical of full random matrices, such as the power-law decay $\propto t^{-3}$ and a deep correlation hole.

\begin{acknowledgments}
E.J.T.-H. and J.A.M.-B. acknowledge funding from VIEP-BUAP (Grant Nos.~MEBJ-EXC19-G, LUAG-EXC19-G, and CELU-EXC19-I), 
Mexico. They are also grateful to LNS-BUAP for allowing the use of the supercomputing facility. 
L.F.S. was supported by the NSF Grant No.~DMR-1603418. 
\end{acknowledgments}

\appendix
\section{Derivation of the expressions for the local density of states and survival probability for the 1D tight-binding model with $W=0$.}
\label{sec:derivation}
Here, we show the steps to obtain Eq.~\eqref{eq:LDOSAn} and Eq.~\eqref{eq:SPno} of the main text. Let us start with the LDOS defined in Eq.~(\ref{Eq:LDOS}).
\subsection{Local density of states} 
When $W=0$, using Eq.~\eqref{Eq:psiAlpha} for the eigenstates, we can write the LDOS as
\begin{eqnarray*}
\rho_{n_0}&&(E)= \frac{2}{L+1}  \sum_{\alpha=1}^L \sin^2 \left( \frac{\alpha n_0 \pi}{L+1} \right) \delta(E-E_{\alpha})\nonumber\\
&&=\frac{2}{L+1} \sum_{\alpha}  \sin^2 \left( \frac{\alpha n_0  \pi}{L+1} \right) \int_{-\infty}^{\infty} \frac{d\tau}{2\pi} e^{i \tau [E-2\cos \left(\alpha \pi/(L+1) \right)]}. \nonumber 
\end{eqnarray*}
In the continuum, $\alpha \pi/(L+1) \rightarrow q$ and 
\[
\dfrac{1}{L} \sum_\alpha \rightarrow \int_{0}^{2\pi} \dfrac{dq}{2\pi},
\]
so
\begin{equation}
\rho_{n_0}(E) \approx\int_{-\infty}^{\infty} \frac{d\tau}{\pi} e^{i \tau E}  \int_{0}^{2\pi} \frac{ dq}{2\pi} e^{ -i2\tau\cos q}  \sin^2(n_0 q).
\label{Eq:rhonoA}
\end{equation}
Using Euler's formula, the second integral in the equation above is separated in three terms,
\ba
&&\int_{0}^{2\pi} \frac{ dq}{2\pi} e^{ -i2\tau\cos q}  \sin^2(n_0 q)= \nonumber \\
&-& \frac{1}{4} \left[ \int_{0}^{2\pi} \frac{ dq}{2\pi} e^{ -i(2\tau\cos q-2n_0 q)} +\int_{0}^{2\pi} \frac{ dq}{2\pi} e^{ -i(2\tau\cos q+2n_0 q)}\right. \nonumber \\
&-& \left. 2\int_{0}^{2\pi} \frac{ dq}{2\pi} e^{ -i 2\tau\cos q}  \right]. \nonumber \\
\label{Eq:IntA}
\ea
To solve the integrals, we use the definition of the Bessel function of first kind,
\be
{\cal J}_n(\gamma)=\int_{0}^{2\pi}\frac{ du}{2\pi} e^{ i(\gamma\sin u - nu)} ,
\label{eq:Bessel}
\ee
and
\be
i^n{\cal J}_n(\gamma)=\int_{0}^{2\pi}\frac{ du}{2\pi} e^{ i(\gamma\cos u - nu)} .
\label{eq:Bessel2}
\ee
This gives us
 \be
 \int_{0}^{2\pi} \frac{ dq}{2\pi} e^{ -i2\tau\cos q}  \sin^2(n_0 q)=\frac{1}{2}\left[{\cal J}_0 (2 t)-(i)^{2n_0}{\cal J}_{2n_0} (2 t)\right] .
 \label{Eq:SPAmp}
 \ee
 
To obtain the LDOS, we still need to compute the integral over $\tau$, that is
\be
\rho_{n_0}(E)=\int_{-\infty}^{\infty} \frac{d\tau}{2\pi} e^{i \tau E}\left[{\cal J}_0 (2 t)-(i)^{2n_0}{\cal J}_{2n_0} (2 t)\right],
\label{Eq:LDOSpre}
\ee 
which leads to 
\be
\rho_{n_0}(E)=\frac{1-\cos\left[2n_0\arccos\left(\frac{E}{2}\right)\right]}{\pi\sqrt{4-E^2}}.
\label{Eq:LDOSpre1}
\ee
We recognize in this last expression the definition of the Chebyshev polynomials of the first kind
\begin{equation*}
T_n(u)=\cos(n\arccos u).
\end{equation*}
Thus, we finally obtain Eq.~\eqref{eq:LDOSAn} of the main text
\be
\rho_{n_0}(E)=\frac{1-T_{2n_0}\left(\frac{E}{2}\right)}{\pi\sqrt{4-E^2}}.
\label{Eq:LDOSAnA}
\ee

Notice that as $n_0$ increases, the term ${\cal J}_{2n_0} (2 t)$ in Eq.~(\ref{Eq:LDOSpre}) becomes negligible for an increasingly large time interval. This can be seen from the definition of the Bessel function of the first kind in terms of the Gamma function $\Gamma$,
\[
{\cal J}_n (\gamma) = \sum_{m=0}^{\infty} \frac{(-1)^m}{m! \Gamma (n + m +1)} \left( \frac{\gamma}{2} \right)^{n + 2m} .
\]
For very large $n$, the function ${\cal J}_n (\gamma)$ is only relevant for very large $\gamma$. This implies that
\be
\rho_{n_{0 \gg 1}}(E) \approx \int_{-\infty}^{\infty} \frac{d\tau}{2\pi} e^{i \tau E} {\cal J}_0 (2 t)
=\frac{1}{\pi\sqrt{4-E^2}} ,
\ee
which is the Eq.~(\ref{Eq:recover}) in the main text.

\subsection{Survival Probability}

Recall that the LDOS and the survival probability are connected by the Fourier transform,
\begin{equation}
S_{P_{n_0}} (t) =  \left| \int_{-2}^{2}\rho_{n_0}(E) e^{-iEt} dE \right|^2 .
\end{equation}
Therefore, based on Eq.~(\ref{Eq:rhonoA}), the survival probability amplitude, 
\[
A(t) = \langle \Psi(0) | e^{-iHt} | \Psi(0) \rangle ,
\]
is twice the result in Eq.~(\ref{Eq:SPAmp}),
\begin{equation}
A(t) = {\cal J}_0 (2 t)-(i)^{2n_0}{\cal J}_{2n_0} (2 t),
\end{equation}
so 
\begin{equation}
S_{P_{n_0}} (t) = \left[ {\cal J}_0 (2 t)-(-1)^{n_0}{\cal J}_{2n_0} (2 t) \right]^2,
\label{eq:fim}
\end{equation}
which is the Eq.~(\ref{eq:SPno}) in the main text. 

The result above can also be obtained directly from the Chebyshev polynomials,
\be
S_{P_{n_0}}(t)=\left|\int_{-2}^{2}\frac{1-T_{2n_0}\left(\frac{E}{2}\right)}{\pi\sqrt{4-E^2}}e^{-iEt}dE\right|^2 .
\label{Eq:SPA}
\ee
The first term in the integral above gives 
\be
\int_{-2}^{2}\frac{1}{\pi\sqrt{4-E^2}}e^{-iEt}dE={\cal J}_0(2t).
\ee
The second term is computed as follows
\ba
&&\int_{-2}^{2}\frac{T_{2n_0}\left(\frac{E}{2}\right)}{\pi\sqrt{4-E^2}}e^{-iEt}dE =\int_{-2}^{2}\frac{T_{2n_0}(\frac{E}{2})}{\pi\sqrt{4-E^2}}\cos{(Et)}dE\nonumber\\
&-&i\int_{-2}^{2}\frac{T_{2n_0}\left(\frac{E}{2}\right)}{\pi\sqrt{4-E^2}}\sin{(Et)}dE.
\ea
Because $T_n(-u)=(-1)^nT_n(u)$ and cosine (sine) is an even (odd) function, the second integral vanishes and the first one is
\be
\frac{2}{\pi}\int_{0}^{1}\frac{T_{2n_0}(u)}{\sqrt{1-u^2}}\cos{(2ut)}du=(-1)^{n_0}{\cal J}_{2n_0}(2t),
\ee
where we used $u=E/2$. Thus, we recover Eq.~(\ref{eq:fim}) above.


\begin{thebibliography}{70}
\bibitem [{foo()}]{footnoteSorathia}%
  \BibitemOpen
  \href@noop {} {}\bibinfo {note} {The slope that we obtain, close to 1,
  differs from that found in Ref.~\cite{Sorathia2012}, where the slope was
  slightly larger than 2. This is because in~\cite{Sorathia2012}, they added a
  factor 2 to the parameter $\xi$ that they use.}\BibitemShut {Stop}%
\end{thebibliography}
\end{document}